\documentclass[conference, a4paper]{IEEEtran}
\IEEEoverridecommandlockouts  
\usepackage[table]{xcolor}                            
\usepackage{units}
\usepackage{graphicx}
\usepackage{multirow, makecell}
\usepackage{amsmath,amssymb,latexsym}
\usepackage{rotating}
\usepackage{lettrine}
\usepackage{textcomp}
\usepackage{bm}

\usepackage{hyperref}
\usepackage{lipsum}%
\usepackage[pscoord]{eso-pic}
\newcommand{\placetextbox}[3]{%
  \setbox0=\hbox{#3}
  \AddToShipoutPictureFG*{
    \put(\LenToUnit{#1\paperwidth},\LenToUnit{#2\paperheight}){\vtop{{\null}\makebox[0pt][c]{#3}}}%
  }%
}%

\newcommand{\Bg}{\cellcolor[HTML]{C0F8C0}}
\newcommand{\Br}{\cellcolor[HTML]{E0C080}}
\newcommand{\By}{\cellcolor[HTML]{F0F080}}

\newcommand{\Ex}{\mathrm{E}}
\newcommand{\Var}{\mathrm{Var}}

\title{On the Accuracy and Precision of Moving Averages to Estimate Wi-Fi Link Quality
\thanks{This work has been partially funded by SoBigData.it ``SoBigData.it receives funding from European Union – NextGenerationEU – National Recovery and Resilience Plan (Piano Nazionale di Ripresa e Resilienza, PNRR) – Project: “SoBigData.it – Strengthening the Italian RI for Social Mining and Big Data Analytics” – Prot. IR0000013 – Avviso n. 3264 del 28/12/2021.''}
}

\author{
    \IEEEauthorblockN{
    Gianluca Cena\IEEEauthorrefmark{1},
    Gabriele Formis\IEEEauthorrefmark{1}\IEEEauthorrefmark{2},
    Matteo Rosani\IEEEauthorrefmark{1}\IEEEauthorrefmark{2},
    Stefano Scanzio\IEEEauthorrefmark{1},
    }  
    \IEEEauthorblockA{\IEEEauthorrefmark{1}National Research Council of Italy (CNR--IEIIT), Italy. \IEEEauthorrefmark{2}Politecnico di Torino, Italy.}    
    Email:  \{gianluca.cena, stefano.scanzio\}@cnr.it, \{gabriele.formis, matteo.rosani\}@polito.it
}

\begin{document}
\placetextbox{0.5}{1}{This is the author's version of an article that has been published.}
\placetextbox{0.5}{0.985}{Changes were made to this version by the publisher prior to publication.}
\placetextbox{0.5}{0.97}{The final version of record is available at \href{https://doi.org/10.1109/ETFA61755.2024.10710784}{https://doi.org/10.1109/ETFA61755.2024.10710784}}%
\placetextbox{0.5}{0.05}{Copyright (c) 2024 IEEE. Personal use is permitted.}
\placetextbox{0.5}{0.035}{For any other purposes, permission must be obtained from the IEEE by emailing pubs-permissions@ieee.org.}%

\maketitle
\thispagestyle{empty}
\pagestyle{empty}

\begin{abstract}
The radio spectrum is characterized by a noticeable variability, which impairs performance and determinism of every wireless communication technology.
To counteract this aspect, mechanisms like Minstrel are customarily employed in real \mbox{Wi-Fi} devices, and the adoption of machine learning  for optimization is envisaged in next-generation Wi-Fi 8.
All these approaches require communication quality to be monitored at runtime.

In this paper, the effectiveness of simple techniques based on moving averages to estimate wireless link quality is analyzed, to assess their advantages and weaknesses.
Results can be used, e.g., as a baseline when studying how artificial intelligence can be employed to mitigate unpredictability of wireless networks by providing reliable estimates about current spectrum conditions.
\end{abstract}


\section{Introduction}
\label{sec:introduction}

Wired links, like those employed in switched Ethernet, are characterized by highly-deterministic behavior.
This means that the likelihood 
that frame transmission over a cable fails
is extremely small and can be neglected in most practical situations.
Conversely, this is untrue for wireless networks like, e.g., Wi-Fi.
Although the throughput featured by this kind of networks has increased steadily over the past two decades,
to the point that it is now comparable to Gigabit Ethernet (including 2.5, 5, and 10GBASE-T), 
the same improvements were not achieved for reliability.
When a single transmission attempt is taken into account, i.e., when operating at the physical (PHY) layer, 
there are non-negligible chances that the frame is corrupted while traveling on air,
due to disturbance that impacts on signal propagation
(either electromagnetic noise from power electronics or interference from nearby wireless nodes).
What is worse, the failure probability is likely to vary over time, sometimes abruptly and noticeably.
Automatic repeat request (ARQ) mechanisms implemented at the medium access control (MAC) layer provide data-link users a satisfactory degree of reliability, but the price to be paid is that latency and jitters may worsen consistently due to both the variable number of retries and random exponential backoff.

When distributed industrial applications are considered, characterized by both reliability and responsiveness constraints,
where devices are interconnected by means of wireless links (to support, e.g., mobility),
suitable metrics are required for expressing the quality of communication at runtime.
Should the quality of links fall below given thresholds, countermeasures must be taken to prevent system malfunctions, damages to equipment, and injuries to human operators.
For example, when fleets of cooperating Autonomous Mobile Robots (AMR) are involved, they could be slowed down and possibly stopped altogether if proper coordination is no longer guaranteed due to a (temporary) communication outage.

Besides reliability, customarily quantified using the packet loss ratio (PLR), the most relevant metric of time-aware applications, like distributed control systems in industrial plants, is latency.
Often, the deadline miss ratio (DMR) is used to describe how many packets arrive to destination too late.
More in general, knowing (at any given time) the complementary cumulative distribution function (CCDF) of the communication latency permits to evaluate the probability with which timing constraints are expected to be violated. 
From this point of view, lost messages and severely delayed ones are treated as equivalent.

In this paper, we follow a different (and much simpler) approach, by considering the \textit{failure probability} $\epsilon$ for single transmission attempts over a wireless link.
The value of $\epsilon$ is generally uninteresting to applications, since they rely on the reliable MAC transmission services made available by the data-link layer.
Nevertheless, it permits, in theory, to infer indirectly all the quantities of interest for them,
including the PLR (that coincides with the probability that all the retries for a given packet fail) 
as well as statistics on communication latency and power consumption (both of which depend on the number of retries actually performed, which cannot exceed the retry limit).

The above model can be characterized by means of measurements performed on commercial devices deployed in a real environment.
For example, the statistical frequency with which transmission attempts fail can be used to infer model parameter~$\epsilon$.
Unfortunately, spectrum conditions are not stationary, which leads to the need to resort to techniques like moving averages (MA).
As is well known, MA usage implies some trade-off: 
using few samples enlarges
estimation jitters due to 
their randomness, worsening precision,
whereas 
increasing their number (and hence the time window on which they are acquired)
makes the estimation procedure unable to promptly track variations of disturbance, ultimately resulting in poor accuracy.
For this reason, the need arises to find an optimal compromise. 
To this extent, in the following we will consider the mean squared error (MSE) with which the instantaneous value of the failure 
rate
is estimated by comparing several datasets, both generated artificially and acquired from experiments.

The paper is structured as follows:
in Section~\ref{sec:link} the problem of link quality estimation is introduced and a simple model is derived that assumes that transmission attempts are statistically independent, 
whereas in Section~\ref{sec:stat} some equations are given that provide the precision offered by two approaches based on moving averages in the case the spectrum is stationary.
Section~\ref{sec:nonstat} analyzes the case of non-stationary conditions, and checks under what conditions the previously derived expressions can be applied satisfactorily.
Some conclusions are finally drawn in Section~\ref{sec:conc}.

\section{Wireless Link Quality Estimation}
\label{sec:link}
The ability to satisfactorily characterize the behavior of wireless links, in spite of their intrinsic randomness, is the core of several techniques aimed to optimize throughput and determinism of next-generation high performance wireless networks.
For this reason, the adoption of machine learning (ML) techniques is explicitly envisaged in Wi-Fi 8 \cite{giordano2023wifi}, as well as in 6G cellular networks \cite{2023-IEEE-Acc-6G, 2023-IEEE-Comm-Sur}.
We distinguish between link quality estimation and prediction: 
the former refers to the ability to provide at runtime reliable estimates of the current instantaneous quality, whereas the latter aims to foresee what will happen in the near future.
In this paper we 
only deal with estimation, as it is intrinsically simpler than prediction.
It is worth noting that estimation can be trivially seen as the limit case of prediction.

Several works in the literature propose using ML to model some aspects that impact on wireless network behavior \cite{ACC-2022, TCCN-2022, JSYST-2021}.
Some of them focus explicitly on Wi-Fi: this is the case of
\cite{IEEE-ComSur-2022-Szott}, \cite{TMC-2021-CogWiFi}, and \cite{Short-Term-Prediction}.
One of the problems of ML-based algorithms (e.g., the different flavors of neural networks) is that their operations are very complex and their performance heavily depends on training.
Conversely, results provided by simpler techniques, like moving averages and regression,
are much easier to understand and explain \cite{2023-WFCS-Formis}, \cite{INDIN-2023-Formis}.
Using the latter as benchmarks against which to check the performance of the former could reveal very useful to understand the real potential of ML applied to wireless spectrum modeling.
This was preliminarily done, to some extent, in \cite{2022-ITL-ML}.

Here, we move one step further in that direction, by considering two techniques for estimating the failure probability of a link that rely on low-pass filters based on moving averages.
In particular, a very simple model is presented that describes the estimation precision achieved by these techniques, under the assumption that spectrum conditions are stationary.
Then, we evaluated to what extent the closed-form expressions we derived can be applied to real non-stationary conditions.

\subsection{Characterization of a Wi-Fi Link}
In the following the IEEE 802.11 protocol \cite{IEEE802-11-2021} (Wi-Fi) will be explicitly considered.
However, most of what we say applies with minimal changes to other wireless transmission technologies as well.
A real Wi-Fi link can be characterised by repeatedly performing acknowledged \textit{one-shot} frame transmissions (i.e., where retransmissions are disabled), which permit to probe the quality of the underlying channel as seen by the sender.
Every time the frame used for probing is received correctly, the recipient returns an explicit confirmation (ACK frame).
By setting the retry limit of the sender to $0$, no retransmissions are performed when the ACK frame is not heard
and the ACK timeout expires.
Starting from these two mutually exclusive and exhaustive events, namely, ACK receptions and timeouts, the sender can evaluate the link failure rate (statistical frequency).
As shown by the Software-Defined MAC (SDMAC) framework \cite{2019-TII-SDMAC}, slight modifications to the open-source drivers of popular commercial Wi-Fi boards permit to detect and log these events on a per-frame basis.

Let $x_i$ denote the outcome of transmission attempt $i$ (where $i\geq 1$) in a given experiment, either $1$ for success or $0$ for failure.
Every value $x_i$ corresponds to a directly \textit{observable} event in the real system.
The sequence $(x_i)_{i\geq 1}$ can be seen as an instance of a discrete-time binary random process
(whose state space coincides with the set $\{0,1\}$).
Let $\epsilon_i$ be the failure probability of attempt $i$.
Generally speaking, this kind of random process can be modeled by defining the failure probability as a function of time, 
$\epsilon: \mathbb{R}\rightarrow[0,1], t\mapsto\epsilon(t)$.
Admittedly, this could be a rough approximation of reality, especially when attempts are close to each other in time, as statistical dependence among them is neglected.
Nevertheless, this model is typically acceptable provided that attempts are spaced wide enough, so that outcomes are (mostly) uncorrelated.

\begin{figure}
    \centering
    \includegraphics[width=1\columnwidth]{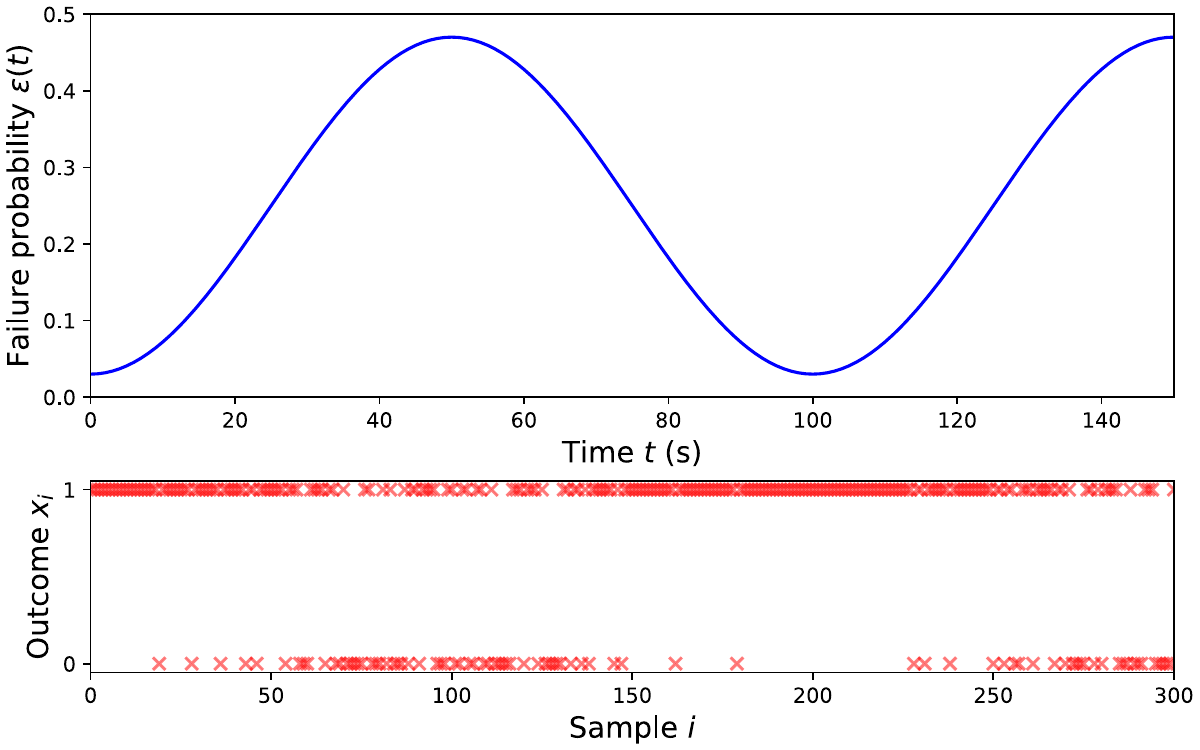}
    \caption{Observable events $x_i$ and hidden model parameters $\epsilon(t)$.}
    \label{fig:model}
\end{figure}

If the time between two subsequent attempts is fixed, as in our experiments (where channel probing is cyclic with period $T_\mathrm{s}$),
we can write $\epsilon_i = \epsilon(i T_\mathrm{s})$, that is, sequence $(\epsilon_i)_{i\geq 1}$ is obtained by ``sampling'' $\epsilon(t)$.
It is worth stressing that $\epsilon(t)$ is just a model parameter, which can not be measured (and not even observed) directly in a real system, and the same applies to the sequence of values $\epsilon_i$.
However, the latter can be estimated from observable events like those described by $x_i$ values.
An example that illustrates this simple link model is depicted in Fig.~\ref{fig:model}, 
where the shape of $\epsilon(t)$ (depicted in the upper part) has both a fixed and a variable (cosine) contribution, while a specific instance of the random process is shown below.
While obtaining $\epsilon_i$ values back from $x_i$ outcomes is generally impossible, computing for them satisfactory estimates is often feasible under specific assumptions.

\subsection{Estimation of Link Quality}
Among the different metrics that can be used to assess the quality of a wireless link, 
in the following we will consider the \textit{frame delivery ratio} (FDR),
defined as the number of frames successfully delivered in a given interval divided by the total number of attempts performed in that interval.
The FDR coincides with the \textit{simple moving average} (SMA) of transmission outcomes.
In particular, at any time step $i$ let 
\begin{align}
    \label{eq:smau}
    u_i= \frac{1}{m}
    \sum_{j=i-m+1}^{i} x_j
\end{align}
denote the FDR evaluated on the most recent $m$ samples, while
\begin{align}
    \label{eq:smav}
    v_i= \frac{1}{m}
    \sum_{j=i+1}^{i+m} x_j
\end{align}
refers to a future interval that includes $m$ samples, and is therefore non-causal. 

The most sensible choice for characterizing the instantaneous quality, we take as the target of our investigation, is
\begin{align}
    \label{eq:smaz}
    z_i = \frac{u_i + v_i}{2} = \frac{1}{2m} \sum_{j=i-m+1}^{i+m} x_j,
\end{align}
which represents the FDR in a \textit{reference interval} $[(i\!-m\!+1) T_\mathrm{s}, (i+m)T_\mathrm{s}]$, centered around the current time step $i$, which includes $2 m$ outcomes.
Unfortunately, this value can be only evaluated ``a posteriori'', since it refers in part to outcomes of future attempts, and hence it cannot be used to optimize network operation at runtime.
Conversely, it represents a good choice as the target for offline training of ML-based prediction models,
e.g., artificial neural networks (ANN) \cite{2023-ACCESS-Colletto}.

If FDR evaluation is required to be carried out at runtime, no samples $x_j$ in the future (for which \mbox{$j>i$}) can be used.
In this case $z_i$ can be approximated by $u_i$ (or, equivalently, $u_i$ can be used to estimate $z_i$).
In theory, a SMA $u'_i$ could be used that spans over an interval (in the past) other than  $[i-m+1,i]$.
Nevertheless, shifting this interval back in the past worsen precision in non-stationary conditions, because it makes spectrum variations between the two intervals larger.
Moreover, 
for space reasons we will not investigate the reason why the width of 
the past
interval has been selected equal to $m$ (half of the reference interval, that we assume to be decided by the user).
Intuitively, this is a trade-off aimed to balance the contributions to the estimation error due to the variability of $v_i$ and $u'_i$.
Therefore, in the following, estimation of the FDR $z_i$ through SMA will rely directly on $u_i$,
$\hat{z}^{\mathrm{SMA}}_i = u_i$.

A possible drawback of the SMA is that, weighting all the previous samples in the same way might not provide the best estimates when channel conditions are not stationary, because of the lag with which estimation is calculated.
As pointed out in \cite{2023-WFCS-Formis}, the \textit{exponential moving average} (EMA) of outcomes $x_i$, defined as
\begin{align}
    \label{eq:ema}
    y_i=\alpha x_i + (1-\alpha) y_{i-1}
\end{align}
could be a more effective way to estimate the current FDR $z_i$ of the link,
i.e., $\hat{z}^{\mathrm{EMA}}_i = y_i$.
Unlike the SMA, it gives more weight to recent samples, and hence it might seemingly suffer less from non-stationarity of the wireless channel. 
Most important, evaluating the EMA demands a very low computation complexity in terms of both CPU and memory usage.
By letting $\beta = 1-\alpha$, \eqref{eq:ema} can be rewritten as
\begin{align}
    \label{eq:emab}
    y_i = \beta y_{i-1} + (1-\beta) x_i = \beta^i y_0 + (1-\beta) \sum_{j=1}^{i} \beta^{i-j} x_j  
\end{align}

To assess how good moving averages are to make estimations about the link FDR,
let $d_i=z_i-u_i$ be the \textit{instantaneous error} between the target $z_i$ and its SMA estimate $u_i$,
while $e_i=z_i-y_i$ refers to the EMA, that is, $y_i$.
A reasonable choice for synthetically describing the precision of SMA and EMA on the whole is their MSE, 
which can be evaluated ``a posteriori'' from experimental outcomes as
\begin{align}
    \mu_{d^2} = \frac{1}{N} \sum_{i=1}^{N} d_i^2, \;\;\;
    \mu_{e^2} = \frac{1}{N} \sum_{i=1}^{N} e_i^2,
\end{align}
where $N$ is the number of samples in the considered dataset
for which both $d_i$ and $e_i$ are defined and can be computed.

\subsection{Probabilistic Model}
Let us model every transmission outcome $x_i$ as a random variable $\mathbf{x}_i$,
which may assume value $0$ with probability $\epsilon_i$ and value $1$ with probability $1-\epsilon_i$.
Hence, the expected value of 
$\mathbf{x}_i$ at discrete time $i$ is $\Ex[\mathbf{x}_i] = 1-\epsilon_i$,
and its probabilistic variance is $\Var(\mathbf{x}_i) = \epsilon_i (1-\epsilon_i)$.
Also 
estimators
$u_i$, $v_i$, $z_i$, and $y_i$, which are computed as linear combinations of outcomes $(x_i)_{i\geq 1}$, can be modeled as random variables, $\mathbf{u}_i$, $\mathbf{v}_i$, $\mathbf{z}_i$, and $\mathbf{y}_i$, respectively.

The expected value of the target FDR $\mathbf{z}_i$ at time $i$, computed according to \eqref{eq:smaz}, which represents the link quality metric we wish to estimate, is
\begin{align}
\label{eq:Bzieps}
    \Ex[\mathbf{z}_i] &= \frac{\Ex[\mathbf{u}_i] + \Ex[\mathbf{v}_i]}{2} 
        = \frac{1}{2m} \sum_{j=i-m+1}^{i+m} \Ex[\mathbf{x}_j] 
        = 1 - \bar\epsilon_{[i]}.
\end{align}
and coincides with one minus the
mean 
value
$\bar\epsilon_{[i]}$ of the failure probability in the $i$-th reference interval.
Eq.~\eqref{eq:Bzieps} relates the expected value at a specific time of a quantity derived from random variables that model observable events (the attempts) and the arithmetic mean of the model parameters $\epsilon_j$ over an interval centered around that time.
A large part of the papers in the literature that focus on the evaluation of link quality in terms of the FDR rely on this well-known reasoning.

The expected value of the SMA estimate provided by \eqref{eq:smau} is 
\begin{align}
    \label{eq:Eui}
    \Ex[\mathbf{u}_i] &= \frac{1}{m} \sum_{j=i-m+1}^{i} \Ex[\mathbf{x}_j],
\end{align}
whereas that provided by the EMA in \eqref{eq:emab} is 
\begin{align}
    \label{eq:Eyi}
    \Ex[\mathbf{y}_i] = \beta^{i} \Ex[\mathbf{y}_0] + (1-\beta) \sum_{j=1}^{i} \beta^{i-j} \Ex[\mathbf{x}_j].
\end{align}

Probabilistic variance can be easily computed under the assumption that random variables $\mathbf{x}_i$ are uncorrelated.
This is not unrealistic, provided that transmission attempts used for probing the link are spaced adequately (as in the experiments we did, where we set $T_\mathrm{s}=\unit[0.5]{s}$, corresponding to a slow $\unit[2]{Hz}$ probing rate).
The variance 
for $\mathbf{u}_i$ is
\begin{align}
    \label{eq:varUi}
    \Var(\mathbf{u}_i) = \frac{1}{m^2}\!\sum_{j=i-m+1}^{i}\!\Var(\mathbf{x}_j),
\end{align}
for $\mathbf{z}_i$ it is 
\begin{align}
    \Var(\mathbf{z}_i) = \frac{\Var(\mathbf{u}_i)\!+\!\Var(\mathbf{v}_i)}{4} 
        = \frac{1}{4m^2}\!\sum_{j=i-m+1}^{i+m}\!\Var(\mathbf{x}_j),
\end{align}
while for $\mathbf{y}_i$ it is
\begin{align}
    \label{eq:Vyi}
    \Var(\mathbf{y}_i) = \beta^{2i} \Var(\mathbf{y}_0) + (1-\beta)^2 \sum_{j=1}^{i} \beta^{2(i-j)} \Var(\mathbf{x}_j)
\end{align}

Variance of $\mathbf{z}_i$ can be rewritten as 
\begin{align}
    \label{eq:varZdis}
    \Var(\mathbf{z}_i) =& 
    \frac{1}{4m^2} \sum_{j=i-m+1}^{i+m} \epsilon_j (1-\epsilon_j) = \\
    =& \frac{1}{2m} \left[ \frac{1}{2m} \sum_{j=i-m+1}^{i+m} \epsilon_j 
    -\frac{1}{2m} \sum_{j=i-m+1}^{i+m} \epsilon_j^2 \right] \nonumber \\
    =& \frac{1}{2m} \left[ \bar \epsilon_{[i]} - {\bar\epsilon_{[i]}}^2 -s^2_{\epsilon_{[i]}} \right] =
    \frac{\bar\epsilon_{[i]} ( 1-\bar\epsilon_{[i]} )}{2m} - \frac{s^2_{\epsilon_{[i]}}}{2m} \nonumber
\end{align}
where $s^2_{\epsilon_{[i]}}$ is the uncorrected variance of $\epsilon_j$ values in the  $i$-th reference interval.
As can be seen, $\Var(\mathbf{z}_i)$ is always less than the case where the failure probability $\epsilon_j$ is constant over the whole interval, in which case $\epsilon_j = \bar\epsilon_{[i]}, \forall j \in [i-m+1, i+m]$ and $s^2_{\epsilon_{[i]}} = 0$.
Unlikely this property is useful for FDR estimation, as $\epsilon_i$ values are unknown and constitute what we wish to determine from transmission outcomes. 

Also estimation errors $d_i$ and $e_i$ can be modeled as random variables, 
$\mathbf{d}_i=\mathbf{z}_i-\mathbf{u}_i$ and $\mathbf{e}_i=\mathbf{z}_i-\mathbf{y}_i$, respectively.
In the following we will check whether or not their probabilistic variance, that is, 
$\Var(\mathbf{d}_i)$ and $\Var(\mathbf{e}_i)$, can be used to satisfactorily estimate the related MSE.

\section{Stationary Conditions}
\label{sec:stat}
Let us initially assume that the random process described by the sequence $(\mathbf{x}_i)_{i\geq 1}$ is stationary.
This means that every random variable $\mathbf{x}_i$ does not depend on $i$, and can be described simply as $\mathbf{x}$,
whose failure probability is $\epsilon$.
The related expected value and variance will be denoted $\Ex[\mathbf{x}]=1-\epsilon$ and $\Var(\mathbf{x})=\epsilon(1-\epsilon)$, respectively.

The same holds for the other random variables that derive from $\mathbf{x}$.
The expected value of the link quality $\mathbf{z}$ computed ``a posteriori'' is
\begin{align}
    \Ex[\mathbf{z}] &= 
    \frac{1}{2m} \sum_{j=i-m+1}^{i+m} \Ex[\mathbf{x}] = \Ex[\mathbf{x}] = 1-\epsilon,
\end{align}
i.e., FDR $\mathbf{z}$ is
an unbiased estimator of the 
success probability $1-\epsilon$ of attempts,
while its probabilistic variance is 
\begin{align}
    \label{eq:varZ}
    \Var(\mathbf{z}) &= 
    \frac{1}{4m^2}\!\sum_{j=i-m+1}^{i+m}\!\Var(\mathbf{x}) = 
    \frac{\Var(\mathbf{x})}{2m} = \frac{\epsilon (1-\epsilon)}{2m},
\end{align}
which, as well known from theory, decreases linearly with the number of samples in the reference interval ($2 m$, in our case).
As expected, neither of them depends on $i$.

Similar reasoning applies to the causal SMA filter $\mathbf{u}$, which produces FDR estimations at runtime,
\begin{align}
    \label{eq:varU}
    \Ex[\mathbf{u}] &= 
    \frac{1}{m} \sum_{j=i-m+1}^{i} \Ex[\mathbf{x}] = \Ex[\mathbf{x}] = 1-\epsilon, \\ 
    \Var(\mathbf{u}) &= 
    \frac{1}{m^2} \sum_{j=i+1}^{i+m} \Var(\mathbf{x}) = 
    \frac{1}{m} \Var(\mathbf{x}) = \frac{\epsilon (1-\epsilon)}{m}.
\end{align}

Concerning the estimate $\mathbf{y}$ provided by the EMA, some time is needed to reach steady-state conditions.
Since the initial value $y_0$ is typically selected as a fixed value, $\Ex[\mathbf{y}_0]=y_0$ and $\Var(\mathbf{y}_0)=0$.
By exploiting stationarity of $\mathbf{x}$, 
\eqref{eq:Eyi} becomes
\begin{align}
    \Ex[\mathbf{y}_i] &= 
    \beta^{i} y_0 + \Ex[\mathbf{x}] (1-\beta) \sum_{j=1}^{i} \beta^{i-j} = \\ \nonumber
    &= \beta^{i} y_0 + \Ex[\mathbf{x}] (1-\beta) \frac{1 - \beta^{i}}{1 - \beta} = \\ \nonumber
    &= \Ex[\mathbf{x}] + \beta^{i} ( y_0 - \Ex[\mathbf{x}] ),  
\end{align}
which has a bias,
while its probabilistic variance given by
\eqref{eq:Vyi} becomes
\begin{align}
    \label{eq:varYi}
    \Var(\mathbf{y}_i) &= 
    \Var(\mathbf{x}) (1-\beta)^2 \sum_{j=1}^{i} \beta^{2(i-j)} = \\ \nonumber
    &= \Var(\mathbf{x}) (1-\beta)^2 \frac{1 - \beta^{2i}}{1 - \beta^{2}} =  \nonumber
    \Var(\mathbf{x}) (1 - \beta^{2i}) \frac{1 - \beta}{1 + \beta}.
\end{align}

Steady-state behavior of $\Ex[\mathbf{y}_i]$ and $\Var(\mathbf{y}_i)$, that is,
when enough time has elapsed so that the initial transient is finished and the EMA has settled, 
can be studied by considering their limits as $i$ approaches infinity.
In particular, since $\beta<1$,
\begin{align}
    \Ex[\mathbf{y}] &= 
    \lim_{i\to\infty} \Ex[\mathbf{y}_i] = \Ex[\mathbf{x}] = 1-\epsilon
\end{align}
which means that, provided enough samples have been fed into the EMA filter, 
$\mathbf{y}$
satisfactorily behaves as an unbiased estimator of the FDR. 
Similarly,
\begin{align}
    \label{eq:varY}
    \Var(\mathbf{y}) &=
    \lim_{i\to\infty} \Var(\mathbf{y}_i) = \frac{1 - \beta}{1 + \beta} \Var(\mathbf{x}) = \\ \nonumber
    &= \frac{\alpha}{2 - \alpha} \Var(\mathbf{x}) 
    = \frac{\alpha}{2 - \alpha} \epsilon (1-\epsilon) 
    \geq \frac{\alpha \epsilon (1-\epsilon)}{2}.
\end{align}

Also processes describing the FDR estimation error are stationary:
in the SMA case it can be described as 
\begin{align}
    \mathbf{d} = \mathbf{z}-\mathbf{u} = \frac{\mathbf{v}+\mathbf{u}}{2}-\mathbf{u}
    = \frac{\mathbf{v}}{2} - \frac{\mathbf{u}}{2}
\end{align}
whereas for the EMA, after the initial transient, it becomes
\begin{align}
    \mathbf{e} = \mathbf{z}-\mathbf{y} = \frac{\mathbf{v}+\mathbf{u}}{2}-\mathbf{y}
    = \frac{\mathbf{v}}{2} - \left( \mathbf{y} - \frac{\mathbf{u}}{2} \right).  
\end{align}

Their expected values are
$\Ex[\mathbf{d}] = \Ex[\mathbf{z}] - \Ex[\mathbf{u}] = 0$
and
$\Ex[\mathbf{e}] = \Ex[\mathbf{z}] - \Ex[\mathbf{y}] = 0$,
which implies that they are accurate estimators.

Concerning variance, we have to distinguish between $\mathbf{y}$ and $\mathbf{u}$, which are calculated on samples in the past interval (and are hence correlated), and $\mathbf{v}$, which refers to the future interval.
The probabilistic variance of the SMA estimation error is
\begin{align}
    \label{eq:varD}
    \Var(\mathbf{d}) 
    = \frac{1}{4} \left( \Var(\mathbf{v}) + \Var(\mathbf{u}) \right)
    = \frac{1}{2m} \Var(\mathbf{x})
\end{align}
Instead, for EMA it is
\begin{align}
    \label{eq:vare}
    \Var(\mathbf{e}) 
    = \Var\left(\frac{\mathbf{v}}{2}\right) + \Var\left(\mathbf{y} - \frac{\mathbf{u}}{2}\right).
\end{align}
To simplify the rightmost term, we start from \eqref{eq:emab} and \eqref{eq:smau} and
split the sums of random variables $\mathbf{x}_j$ in two disjoint intervals,
$[1, i-m]$ and $[i-m+1,i]$, 
\begin{align}
    \mathbf{y}_i - \frac{\mathbf{u}_i}{2} 
    &= \beta^i y_0 + \sum_{j=1}^{i-m} (1-\beta) \beta^{i-j} \mathbf{x}_j + \\ \nonumber
    &+ \sum_{j=i-m+1}^{i} \left[ (1-\beta) \beta^{i-j} - \frac{1}{2 m} \right] \mathbf{x}_j.
\end{align}
Sums refer to different sets of random variables $\mathbf{x}_j$,
so they are uncorrelated and the variance of the sum equals the sum of variances.
Since the channel (i.e., $\mathbf{x}_i$) is assumed to be stationary, from \eqref{eq:varUi} and \eqref{eq:varYi} we have
\begin{align}
    &\Var\left( \mathbf{y}_i - \frac{\mathbf{u}_i}{2} \right) = \Var(\mathbf{x}) \cdot \xi 
\end{align}
where
\begin{align}
    \xi &=  \sum_{j=1}^{i-m} \left[ (1-\beta) \beta^{i-j} \right]^2 +  
    \sum_{j=i-m+1}^{i} \left[ (1-\beta) \beta^{i-j} - \frac{1}{2 m} \right]^2 \nonumber \\
    &= \sum_{j=1}^{i} (1-\beta)^2 \beta^{2(i-j)} + \sum_{j=i-m+1}^{i} \left[ \frac{1}{4 m^2} -  
    \frac{(1-\beta) \beta^{i-j}}{m} \right] \nonumber \\ 
    &= (1 - \beta)^2 \frac{1 - \beta^{2i}}{1 - \beta^2} + \frac{1}{4 m} 
    - \frac{1-\beta}{m} \frac{1-\beta^m}{1-\beta} 
\end{align}
and, when the initial transient has finished
\begin{align}
    &\Var\left(\mathbf{y} - \frac{\mathbf{u}}{2}\right) = 
    \Var(\mathbf{x})\left[ \frac{1 - \beta}{1 + \beta} + \frac{1}{4 m} - \frac{1-\beta^m}{m} \right].
\end{align}

Since $\Var(\mathbf{v}/2) = \Var(\mathbf{x})/4m$, from \ref{eq:vare} we have
\begin{align}
    \label{eq:varE}
    \Var\left(\mathbf{e}\right) &= 
    \Var(\mathbf{x})\left[ \frac{1 - \beta}{1 + \beta} + \frac{1}{m} \left( \beta^m -\frac{1}{2} \right) \right] \\
    &= \Var(\mathbf{x})\left[ \frac{\alpha}{2 - \alpha} + \frac{1}{m} (1-\alpha)^m -\frac{1}{2 m} \right].\nonumber 
\end{align}

From a probabilistic point of view the MSE of the SMA and the EMA can be described as $\Ex[\mathbf{d}^2]$ and $\Ex[\mathbf{e}^2]$, respectively.
In turn, they are equal to $\Ex[\mathbf{d}]^2 + \Var(\mathbf{d})$ and $\Ex[\mathbf{e}]^2 + \Var(\mathbf{e})$.
Since in stationary conditions both $\mathbf{u}$ and $\mathbf{y}$ are unbiased estimators of $\mathbf{z}$, 
$\Ex[\mathbf{d}]^2=0$ and $\Ex[\mathbf{e}]^2=0$,
which means that the MSE corresponds to the probabilistic variance.
However, this is untrue when spectrum conditions vary over time.

\begin{table*}[t]
    \centering
    \caption{Stationary conditions (synthetic data): FDR estimation error stats and probabilistic variance for EMA ($e$) and SMA ($d$)}
    \label{tab:stats_stat2}
    \tabcolsep=0.1cm
    \begin{tabular}{c|rc|ccccc|ccccc}
        Disturbance & \multicolumn{2}{c|}{Filter params} & \multicolumn{5}{c|}{EMA FDR estim. error (mean, var, MSE, prob. var, MAE)} & \multicolumn{5}{c}{SMA FDR estim. error (mean, var, MSE, prob. var, MAE)} \\
        $\epsilon$ & $m$ & $\alpha$ & 
        $\mu_e$ & $\sigma^2_e$   & $\mu_{e^2}$ & $\Var\left(\mathbf{e}\right)$ & $\mu_{|e|}$ &
        $\mu_d$ & $\sigma^2_d$   & $\mu_{d^2}$ & $\Var\left(\mathbf{d}\right)$ & $\mu_{|d|}$ \\
        \hline
\multirow{4}{*}{0.1} &
    10 & 0.2000 &  -7.159744e-009 & 0.006453 & \Bg 0.006453 & \Bg 0.006458 & 0.061112 &  -2.040816e-008 & 0.004491 & \Bg 0.004491 & \Bg 0.004495 & 0.050046   \\ 
&
   100 & 0.0200 &  -2.889396e-008 & 0.000579 & \Bg 0.000579 & \Bg 0.000578 & 0.019158 &  -4.897959e-008 & 0.000449 & \Bg 0.000449 & \Bg 0.000449 & 0.016802   \\ 
&
  1000 & 0.0020 &  +2.328946e-007 & 0.000057 & \Bg 0.000057 & \Bg 0.000057 & 0.006018 &  +1.690816e-007 & 0.000045 & \Bg 0.000045 & \Bg 0.000045 & 0.005336   \\ 
&
 10000 & 0.0002 &  -2.251246e-008 & 0.000006 & \Bg 0.000006 & \Bg 0.000006 & 0.001918 &  -2.965408e-007 & 0.000005 & \Bg 0.000005 & \Bg 0.000004 & 0.001755   \\        
        \hline
\multirow{4}{*}{0.2} &
    10 & 0.2000 &  +4.179911e-009 & 0.011481 & \Bg 0.011481 & \Bg 0.011491 & 0.085507 &  -2.040816e-008 & 0.007991 & \Bg 0.007991 & \Bg 0.007997 & 0.069377   \\ 
&
   100 & 0.0200 &  -3.031618e-007 & 0.001027 & \Bg 0.001027 & \Bg 0.001028 & 0.025564 &  -3.443877e-007 & 0.000798 & \Bg 0.000798 & \Bg 0.000800 & 0.022480   \\ 
&
  1000 & 0.0020 &  -1.029868e-006 & 0.000102 & \Bg 0.000102 & \Bg 0.000102 & 0.008040 &  -1.230765e-006 & 0.000079 & \Bg 0.000079 & \Bg 0.000080 & 0.007108   \\ 
&
 10000 & 0.0002 &  +9.944382e-008 & 0.000010 & \Bg 0.000010 & \Bg 0.000010 & 0.002556 &  +1.261883e-006 & 0.000008 & \Bg 0.000008 & \Bg 0.000008 & 0.002310   \\ 
        \hline
\multirow{4}{*}{0.4} &
    10 & 0.2000 &  -2.768329e-008 & 0.017237 & \Bg 0.017237 & \Bg 0.017242 & 0.106140 &  -6.122448e-008 & 0.012003 & \Bg 0.012003 & \Bg 0.011999 & 0.086243  \\ 
&
   100 & 0.0200 &  -1.478658e-007 & 0.001539 & \Bg 0.001539 & \Bg 0.001542 & 0.031327 &  -2.209183e-007 & 0.001197 & \Bg 0.001197 & \Bg 0.001200 & 0.027574  \\ 
&
  1000 & 0.0020 &  -1.042735e-006 & 0.000154 & \Bg 0.000154 & \Bg 0.000153 & 0.009902 &  -1.125969e-006 & 0.000121 & \Bg 0.000121 & \Bg 0.000120 & 0.008780  \\ 
&
 10000 & 0.0002 &  +3.624603e-007 & 0.000015 & \Bg 0.000015 & \Bg 0.000015 & 0.003134 &  +1.101138e-006 & 0.000012 & \Bg 0.000012 & \Bg 0.000012 & 0.002810  \\ 
    \end{tabular}
\end{table*}

\subsection{Validation}

A numerical evaluation was carried out to double check the correctness of the above analysis and the related mathematical derivations.
Several instances of \textit{synthetic} stationary binary random processes $(x_i)_{i\geq 1}$ were generated using the \texttt{rand()} function of the \texttt{math} C library, each one characterized by a specific failure probability $\epsilon$.
In particular, we selected $\epsilon \in \{0.1, 0.2, 0.4 \}$.
Every process included more than ten million samples, to provide statistically reliable results.
Although the properties of \texttt{rand()} are not optimal in terms of randomness (complete independence between subsequent values is not ensured), it is good enough for our purposes.

Sequences $(z_i)_{i\geq 1}$, $(u_i)_{i\geq 1}$, and $(y_i)_{i\geq 1}$ were then calculated by feeding above outcomes to the related SMA 
and EMA filters.
To properly initialize filters and to tackle the non-causality of $\mathbf{z}$, a prefix and a postfix including $100000$ samples each were prepended/appended to any dataset (generated using the same $\epsilon$), which were not used for computing statistics.
Different configurations were considered for SMA and EMA filters.
To ensure fair comparison, their parameters ($m$ and $\alpha$, respectively) were selected in such a way to make the probabilistic variance of $\mathbf{u}$ and $\mathbf{y}$ about the same.
From \eqref{eq:varU} and \eqref{eq:varY}, and under the hypothesis $\alpha \ll 1$, this means setting $\alpha=2/m$.
In particular, for the SMA we selected $m \in \{10, 100, 1000, 10000 \}$, which means setting the reference interval to $\unit[5]{s}$, $\unit[50]{s}$, $\unit[\sim\!8]{min}$, and $\unit[\sim\!83]{min}$.

Finally, instantaneous FDR estimation errors, as given by sequences $(d_i)$ and $(e_i)$, were computed and statistics on them evaluated.
Results are reported in Table~\ref{tab:stats_stat2}.
The mean signed errors ($\mu_d$ and $\mu_e$) provide an indication about \textit{estimation accuracy}, that is, the presence of systematic errors (bias).
The very small values observed for them confirm that, passed the initial transient, moving averages 
$u_i$ and $y_i$
provide accurate, unbiased estimates of the FDR $z_i$.
Conversely, the MSE committed by SMA and EMA for FDR estimation
coincides with $\mu_{d^2}$ and $\mu_{e^2}$, respectively.
Since process $(x_i)$ is stationary and estimators are unbiased, these values are practically the same as variances $\sigma^2_d$ and $\sigma^2_e$.
For completeness, also the mean absolute errors (MAE) $\mu_{|d|}$ and $\mu_{|e|}$ have been reported.
Both MSE and MAE permit to assess \textit{estimation precision}.

In the table, the probabilistic variance, computed analytically by means of 
\eqref{eq:varD} and \eqref{eq:varE}, is included next to the MSE calculated on samples.
As can be seen, 
$\mu_{d^2} \simeq \Var(\bm{d})$ and $\mu_{e^2} \simeq \Var(\bm{e})$, 
these two quantities are very similar for both the SMA and the EMA, that is,
which confirms correctness of our simple equations.
Results show that, in stationary conditions, closed-form expressions for 
the probabilistic variance
permit to approximate very well the MSE committed by moving averages:
as expected from theory, the MSE is inversely proportional to $m$ for the SMA and grows linearly with $\alpha$ for the EMA.

\section{Non-stationary Conditions}
\label{sec:nonstat}
Above analysis was repeated by considering non-stationary conditions,
where the failure probability is not fixed.
Two cases were taken into account, where the sequence of transmission outcomes $(x_i)$ was either obtained synthetically or derived from experimental results.

\begin{table*}[t]
    \centering
    \caption{Non-stationary conditions (synthetic data): FDR estimation error stats and prob. variance for EMA ($e$) and SMA ($d$)}
    \label{tab:stats_synt2}
    \tabcolsep=0.1cm
    \begin{tabular}{cc|rc|ccccc|ccccc}
        \multicolumn{2}{c|}{Disturbance} & \multicolumn{2}{c|}{Filter params} & \multicolumn{5}{c|}{EMA FDR estim. error (mean, var, MSE, prob. var, MAE)} & \multicolumn{5}{c}{SMA FDR estim. error (mean, var, MSE, prob. var, MAE)} \\
        $f$ & $\epsilon$ & $m$ & $\alpha$ & 
        $\mu_e$ & $\sigma^2_e$   & $\mu_{e^2}$ & $\Var\left(\mathbf{e}\right)$ & $\mu_{|e|}$ &
        $\mu_d$ & $\sigma^2_d$   & $\mu_{d^2}$ & $\Var\left(\mathbf{d}\right)$ & $\mu_{|d|}$ \\
        \hline
\multirowcell{8}{\rotatebox[origin=c]{90}{$\unit[0.0001]{Hz}$}}
&\multirowcell{4}{$0.1 \pm 0.005$}  &
        10 & 0.2000 & -7.159744e-009 & 0.006451 & \Bg 0.006451 & \Bg 0.006457 & 0.061084 &  -2.040816e-008 & 0.004490 & \Bg 0.004490 & \Bg 0.004493 & 0.050025   \\ 
    &&
       100 & 0.0200 & -2.887927e-008 & 0.000579 & \Bg 0.000579 & \Bg 0.000578 & 0.019155 &  -4.897959e-008 & 0.000449 & \Bg 0.000449 & \Bg 0.000449 & 0.016804   \\ 
    &&
      1000 & 0.0020 & +1.953063e-007 & 0.000057 & \Bg 0.000057 & \Bg 0.000057 & 0.006034 &  +1.242857e-007 & 0.000045 & \Bg 0.000045 & \Bg 0.000045 & 0.005352   \\ 
    &&
     10000 & 0.0002 & +3.978745e-007 & 0.000010 & \By 0.000010 & \By 0.000006 & 0.002498 &  +2.091377e-007 & 0.000010 & \By 0.000010 & \By 0.000004 & 0.002609   \\ 
    \cline{2-14}
&\multirowcell{4}{$0.1 \pm  0.05$}  &
       10 & 0.2000  & +1.870554e-008 & 0.006354 & \Bg 0.006354 & \Bg 0.006453 & 0.059161 &  +1.530612e-008 & 0.004424 & \Bg 0.004424 & \Bg 0.004491 & 0.048583  \\ 
    &&
       100 & 0.0200 & -1.189478e-007 & 0.000571 & \Bg 0.000571 & \Bg 0.000577 & 0.018763 &  -1.280612e-007 & 0.000444 & \Bg 0.000444 & \Bg 0.000449 & 0.016487  \\ 
    &&
      1000 & 0.0020 & -2.741337e-007 & 0.000086 & \By 0.000086 & \By 0.000057 & 0.007420 &  -3.341836e-007 & 0.000075 & \By 0.000075 & \By 0.000045 & 0.006934  \\ 
    &&
     10000 & 0.0002 & -2.978835e-007 & 0.000367 & \Br 0.000367 & \Br 0.000006 & 0.017170 &  -2.761887e-007 & 0.000512 & \Br 0.000512 & \Br 0.000004 & 0.020334  \\      
        \hline
\multirowcell{8}{\rotatebox[origin=c]{90}{$\unit[0.001]{Hz}$}}
&\multirowcell{4}{$0.1 \pm 0.005$}  &
        10 & 0.2000 & -7.159744e-009 & 0.006453 & \Bg 0.006453 & \Bg 0.006459 & 0.061103 &  -2.040816e-008 & 0.004491 & \Bg 0.004491 & \Bg 0.004495 & 0.050038   \\ 
    &&
       100 & 0.0200 & -2.890216e-008 & 0.000578 & \Bg 0.000578 & \Bg 0.000578 & 0.019151 &  -4.897959e-008 & 0.000449 & \Bg 0.000449 & \Bg 0.000450 & 0.016806  \\ 
    &&
      1000 & 0.0020 & +1.686608e-007 & 0.000060 & \Bg 0.000060 & \Bg 0.000057 & 0.006197 &  +7.086734e-008 & 0.000050 & \Bg 0.000050 & \Bg 0.000045 & 0.005618   \\ 
    &&
     10000 & 0.0002 & +2.012677e-007 & 0.000006 & \Bg 0.000006 & \Bg 0.000006 & 0.001924 &  -1.124949e-007 & 0.000005 & \Bg 0.000005 & \Bg 0.000004 & 0.001743   \\ 
    \cline{2-14}
&\multirowcell{4}{$0.1 \pm 0.05$}  & 
        10 & 0.2000 & +1.870554e-008 & 0.006360 & \Bg 0.006360 & \Bg 0.006457 & 0.059235 &  +1.530612e-008 & 0.004422 & \Bg 0.004422 & \Bg 0.004494 & 0.048586  \\ 
    &&
       100 & 0.0200 & -1.282116e-007 & 0.000597 & \Bg 0.000597 & \Bg 0.000578 & 0.019207 &  -1.403061e-007 & 0.000471 & \Bg 0.000471 & \Bg 0.000449 & 0.016999  \\ 
    &&
      1000 & 0.0020 & -3.321854e-007 & 0.000413 & \Br 0.000413 & \Br 0.000057 & 0.017705 &  -3.509694e-007 & 0.000546 & \Br 0.000546 & \Br 0.000045 & 0.020616  \\ 
    &&
     10000 & 0.0002 & +5.504237e-007 & 0.000011 & \By 0.000011 & \By 0.000006 & 0.002643 &  +4.384949e-007 & 0.000004 & \Bg 0.000004 & \Bg 0.000004 & 0.001659   \\ 
    \end{tabular}
\end{table*}

\subsection{Synthetic dataset}
In this case, the instances of the random processes $(x_i)$ were generated assuming that the failure probability varies over time according to the law
\begin{align}
    \epsilon(t) = \epsilon_0 + \Delta_\epsilon \cdot \cos{ 2 \pi f t },
\end{align}
which implies
\begin{align}
    \epsilon_i = \epsilon_0 + \Delta_\epsilon \cdot \cos{ 2 \pi f T_\mathrm{s} i}.
\end{align}

Background disturbance, modeled by $\epsilon_0$, was set equal to $0.1$.
According to our previous experience with wireless communications, this is a realistic value in typical conditions (every transmission attempt has $10\%$ probability to fail).
Two values were selected for $\Delta_\epsilon$, which defines the range in which the failure probability may vary
($\epsilon_0 - \Delta_{\epsilon} \leq \epsilon(t) \leq \epsilon_0 + \Delta_{\epsilon}$), 
that is, $\Delta_\epsilon \in \{0.05, 0.005\}$, the second one resembling quasi-stationary conditions.
Frequency $f$ defines instead the rate at which disturbance varies.
We selected $f\in \{\unit[0.001]{Hz}, \unit[0.0001]{Hz}\}$, which correspond to a periodicity of disturbance in the order of $17$ minutes and about three hours, respectively.

Results are reported in Table~\ref{tab:stats_synt2},
where a color code (green, yellow, red) has been employed to highlight to what extent approximating the MSE with the probabilistic variance computed according to the equations in the previous section is acceptable.
The first and third block of rows 
($\Delta_\epsilon = 0.005$) model quasi-stationary processes, where the failure rate varies to a very limited extent.
As can be seen, the MSE in this case is approximated satisfactorily even for quite slow filters ($m \leq 1000$, $\alpha \geq 0.002$).
When the cut-off frequency of SMA and EMA low-pass filters comes close to $f$, approximation becomes poorer but still acceptable (see the case $f=\unit[0.0001]{HZ}$, $m = 10000$, $\alpha = 0.0002$), 

In the second and fourth block of rows, variations of the failure rate experience a tenfold increase, with $\epsilon$ that lies in the range $[0.05,0.15]$.
In this case, approximating the MSE with the probabilistic variance is acceptable only if the cut-off frequency remains about one decade below $f$.
Very large discrepancies are observed, for example, in two rows, corresponding to the cases 
$f=\unit[0.0001]{Hz}$, $m = 10000$, $\alpha = 0.0002$, and 
$f=\unit[0.001]{Hz}$, $m = 1000$, $\alpha = 0.002$.
This is because moving averages configured this way not only are unable to promptly track variations of disturbance,
but are counter-phased with respect to them, which enlarges errors noticeably.
A proof of this can be observed in the fourth block of rows.
By slowing down the filter further (case $f=\unit[0.001]{Hz}$, $m = 10000$, $\alpha = 0.0002$) the MSE diminishes and comes closer to the probabilistic variance.

\begin{table*}[t]
    \centering 
    \caption{Non-stationary conditions (experimental data): FDR estimation error stats and prob. variance for EMA ($e$) and SMA ($d$)}
    \label{tab:stats_exp2}
    \tabcolsep=0.1cm
    \begin{tabular}{c|rc|ccccc|ccccc}
    \;\;\;Disturbance\;\;\; & \multicolumn{2}{c|}{Filter params} & \multicolumn{5}{c|}{EMA FDR estim. error (mean, var, MSE, prob. var, MAE)} & \multicolumn{5}{c}{SMA FDR estim. error (mean, var, MSE, prob. var, MAE)} \\
        $\epsilon$ & $m$ & $\alpha$ & 
        $\mu_e$ & $\sigma^2_e$   & $\mu_{e^2}$ & $\Var\left(\mathbf{e}\right)$ & $\mu_{|e|}$ &
        $\mu_d$ & $\sigma^2_d$   & $\mu_{d^2}$ & $\Var\left(\mathbf{d}\right)$ & $\mu_{|d|}$ \\
        \hline
\multirowcell{4}{Ch 1 \\ $0.071803$}  &
        10 & 0.2000 & +4.969577e-007 & 0.004666 & \Bg 0.004666 & \Bg 0.004789 & 0.047277 &  +5.180195e-007 & 0.003235 & \Bg 0.003235 & \Bg 0.003332 & 0.039206   \\ 
    &    
       100 & 0.0200 & +6.590860e-007 & 0.000430 & \Bg 0.000430 & \Bg 0.000428 & 0.016004 &  +1.036039e-006 & 0.000341 & \Bg 0.000341 & \Bg 0.000333 & 0.014120   \\ 
    &    
      1000 & 0.0020 & -4.227369e-007 & 0.000113 & \By 0.000113 & \By 0.000042 & 0.007023 &  +7.674695e-006 & 0.000127 & \By 0.000127 & \By 0.000033 & 0.006936  \\ 
    &    
     10000 & 0.0002 & +1.507616e-004 & 0.000113 & \Br 0.000113 & \Br 0.000004 & 0.006851 &  +1.124249e-004 & 0.000138 & \Br 0.000138 & \Br 0.000003 & 0.007478  \\ 
    \hline
\multirowcell{4}{Ch 5 \\ $0.063164$}  &
        10 & 0.2000 & +6.532483e-008 & 0.004141 & \Bg 0.004141 & \Bg 0.004252 & 0.043097 &  +2.354634e-007 & 0.002864 & \Bg 0.002864 & \Bg 0.002959 & 0.035841 \\ 
    &    
       100 & 0.0200 & -1.204600e-006 & 0.000376 & \Bg 0.000376 & \Bg 0.000380 & 0.014880 &  -1.469292e-006 & 0.000293 & \Bg 0.000293 & \Bg 0.000296 & 0.013072  \\ 
    &    
      1000 & 0.0020 & -4.619686e-006 & 0.000067 & \By 0.000067 & \By 0.000038 & 0.005779 &  -3.983570e-006 & 0.000067 & \By 0.000067 & \By 0.000030 & 0.005528  \\ 
    &    
     10000 & 0.0002 & -8.130657e-005 & 0.000137 & \Br 0.000137 & \Br 0.000004 & 0.007572 &  -6.934991e-005 & 0.000189 & \Br 0.000189 & \Br 0.000003 & 0.008912  \\ 
    \hline
\multirowcell{4}{Ch 9 \\ $0.14465$}  &
        10 & 0.2000 & -5.726406e-007 & 0.008724 & \Bg 0.008724 & \Bg 0.008890 & 0.072528 &  -4.709269e-007 & 0.006056 & \Bg 0.006056 & \Bg 0.006186 & 0.058984  \\ 
    &    
       100 & 0.0200 & -2.447254e-006 & 0.000815 & \Bg 0.000815 & \Bg 0.000795 & 0.022426 &  -2.924456e-006 & 0.000649 & \Bg 0.000649 & \Bg 0.000619 & 0.019902  \\ 
    &    
      1000 & 0.0020 & +8.169690e-006 & 0.000248 & \By 0.000248 & \By 0.000079 & 0.008900 &  +7.279587e-006 & 0.000281 & \By 0.000281 & \By 0.000062 & 0.008558  \\ 
    &    
     10000 & 0.0002 & +1.801074e-004 & 0.000241 & \Br 0.000241 & \Br 0.000008 & 0.008593 &  +2.087150e-004 & 0.000246 & \Br 0.000246 & \Br 0.000006 & 0.009030  \\
    \hline
\multirowcell{4}{Ch 13 \\ $0.255758$}  &
        10 & 0.2000 & +3.795014e-007 & 0.013374 & \Bg 0.013374 & \Bg 0.013676 & 0.092732 &  +3.767415e-007 & 0.009287 & \Bg 0.009287 & \Bg 0.009517 & 0.075160  \\ 
    &    
       100 & 0.0200 & +2.563495e-006 & 0.001223 & \Bg 0.001223 & \Bg 0.001223 & 0.027746 &  +3.838054e-006 & 0.000978 & \Bg 0.000978 & \Bg 0.000952 & 0.024638  \\ 
    &    
      1000 & 0.0020 & -1.434795e-005 & 0.000352 & \By 0.000352 & \By 0.000121 & 0.011134 &  -1.921711e-005 & 0.000395 & \By 0.000395 & \By 0.000095 & 0.010609  \\ 
    &    
     10000 & 0.0002 & -1.672247e-004 & 0.000320 & \Br 0.000320 & \Br 0.000012 & 0.010295 &  -1.732733e-004 & 0.000326 & \Br 0.000326 & \Br 0.000010 & 0.011162  \\         
    \end{tabular}
\end{table*}

\subsection{Experimental dataset}
Our final analysis is carried out on experimental logs.
In particular, four random processes $(x_i)$ are involved that were acquired from a real testbed that includes commercial devices communicating over Wi-Fi links, as described in Section~\ref{sec:link}.
The testbed was provided with multiple stations (STA) and access points (AP), so that four links tuned on non-overlapping channels 1, 5, 9, and 13 were operated contextually.
Although the equipment we employed for every channel was basically equivalent (Wi-Fi boards were exactly the same and associated to two pairs of identical APs), deployed in the same lab, with transmitter and receiver antennas similarly spaced (about $\unit[3]{m}$), the acquired logs highlighted sensible differences among spectrum conditions, because of the different sets of visible interfering devices (number and spatial position), each one with its own traffic (unknown to us).

Every single log included $1261735$ samples, corresponding to slightly more that one full week of continuous operations.
As in the previous cases, the initial and the final $100000$ samples were not included in the statistics (but were feed to the SMA an EMA filters).

Results are reported in Table~\ref{tab:stats_exp2}.
As can be seen from $\mu_d$ and $\mu_e$, accuracy is still very good.
It worsens a bit when a slower filter is employed (higher $m$ or lower $\alpha$), i.e., when the cutoff frequency is decreased.
This is probably due to the fact that the number of experimental samples was one order of magnitude smaller than for synthetic data, which amplifies differences found at the beginning and at the end of the portion of dataset on which statistics are computed.

Concerning precision ($\mu_{d^2}$ and $\mu_{e^2}$, which also in this case practically coincide with the sample variance $\sigma^2_d$ and  $\sigma^2_e$), when the filter is fast ($m \leq 100$, $\alpha \geq 0.02$) estimation is clearly poor, because of the limited number of samples on which arithmetic means are computed.
In this case, approximating the MSE with the probabilistic variance is acceptable.
Slowing down the filter ($m \geq 1000$, $\alpha \leq 0.002$) improves precision, but the MSE can no longer be satisfactorily approximated by the probabilistic variance.
As intuitively expected, when filters are slow (higher $m$ or lower $\alpha$) the EMA behaves slightly better than the SMA (lower MSE).
However, the real advantage of the EMA is its very low computational effort.

A peculiar phenomenon is also observed: in some cases (channels 1 and 5), decreasing the cutoff frequency makes precision worsen (i.e., the MSE increases).
See, e.g., the rows corresponding to $m=1000$ and $m=10000$.
Likely, this behavior also affected the other channels ($9$ and $13$), but for values of $m$ larger than $10000$.
This is not unexpected, as an excessively slow filter is unable to suitably track FDR variations, which consequently results in higher estimation errors.

As show in \cite{2023-ACCESS-Colletto}, above behavior suggests that, when moving averages are exploited for wireless link quality estimation, a compromise should be found in order to maximize precision, for instance by minimizing the mean error.
Similarly to neural networks, this can be done by a proper training phase, which permits to select filter parameters starting from measurements carried out on devices deployed in the real world.

\section{Conclusions}
\label{sec:conc}
Wireless communications are typically deemed unreliable and not deterministic enough for adoption in real-time control systems, like those found at the shop-floor in industrial plants, because 
the radio spectrum suffers from sudden and mostly unexpected phenomena like  
interference and noise that may corrupt ongoing transmissions.
The ability to estimate link quality at runtime is essential to enable adaptive mechanisms operating, e.g., at the MAC layer, whose purpose is to improve overall network behavior.
This kind of techniques are customarily implemented in commercial equipment (for instance, Minstrel permits to dynamically select the optimal modulation and coding scheme in Wi-Fi), but they will become increasingly important in the future to complement machine learning, as foreseen by Wi-Fi 8.

A relevant metric to assess the instantaneous communication quality of a link, which can be easily determined by the transmitting STA for confirmed traffic, is the FDR evaluated on an interval of a given width and centered on the current time.
Doing so requires a non-causal filter (i.e., one that needs to know the future behavior), which is fine for post-analysis (e.g., for training a neural network) but is unfeasible for mechanisms conceived to  operate at runtime.
For this reason, moving averages like the SMA and the EMA are customarily adopted for estimating the FDR, which can be easily parameterized by means of the window width $m$ and the smoothing factor $\alpha$, respectively.
Both these approaches are accurate, since they provide unbiased estimates.
Precision depends instead on filter configuration and actual disturbance.
A convenient way to express it is the MSE, although other quantities can be also used, like the MAE.

In this paper, some simple formulas are derived for the MSE committed by SMA and EMA for FDR estimation in the simplistic case of stationary conditions.
After numerically verifying their correctness, we applied them to two classes of non-stationary conditions.
First, we generated some synthetic datasets that describe random transmission processes, 
where the failure probability is made up of both a fixed and a variable contribution, the latter characterized by a sinusoidal shape.
Then, we considered some logs obtained from experiments that involve a testbed including real Wi-Fi devices.

In both cases, our theoretical model provided acceptable approximations for the MSE as long as SMA and EMA low-pass filters are fast, in which case the main contribution to the estimation error is due to the unavoidable variability of outcomes (which are satisfactorily described by a random process).
When the cut-off frequency of filters is decreased, however, their inability to promptly track variations of disturbance becomes predominant.
As results show, the MSE reaches a plateau, after which it starts growing again.
This implies that, when moving averages are employed for link quality estimation in real devices, a pre-training phase should be carried out to minimize their MSE.

As clearly highlighted in several recent papers, more sophisticate solutions exist, like neural networks and multi-pole filters, which are a much better option than SMA and EMA.
Nevertheless, results reported here, as well as the discussion on MSE approximation based on what is seen in stationary conditions, can be used as the baseline for evaluating the performance of ML-based solutions.
As future work we plan to define some simple model that relies on closed-form yet manageable expressions to approximately describe the MSE in non-stationary condition, in order to provide a suitable benchmark against which to check effectiveness of advanced quality estimation (and prediction) techniques.

\bibliographystyle{IEEEtran}
\bibliography{bibliography}

\end{document}